\documentclass[twocolumn,floatfix,aps,rmp]{revtex4}
\usepackage{graphicx}
\usepackage{amsmath}
\usepackage{amssymb}
\usepackage{bm}
\setcitestyle{numbers,comma,square,compress}
\begin{document}
\title{Search for Majorana fermions in superconductors}
\author{C. W. J. Beenakker}
\affiliation{Instituut-Lorentz, Universiteit Leiden, P.O. Box 9506, 2300 RA Leiden, The Netherlands}
\date{April 2012}
\begin{abstract}
Majorana fermions (particles which are their own antiparticle) may or may not exist in Nature as elementary building blocks, but in condensed matter they can be constructed out of electron and hole excitations. What is needed is a superconductor to hide the charge difference, and a topological (Berry) phase to eliminate the energy difference from zero-point motion. A pair of widely separated Majorana fermions, bound to magnetic or electrostatic defects, has non-Abelian exchange statistics. A qubit encoded in this Majorana pair is expected to have an unusually long coherence time. We discuss strategies to detect Majorana fermions in a topological superconductor, as well as possible applications in a quantum computer. The status of the experimental search is reviewed.
\medskip\\
{\tt scheduled for vol.\ 4 (2013) of Annual Review of Condensed Matter Physics}
\end{abstract}
\maketitle
\tableofcontents

\section{What are they?}
\label{whatarethey}

A Majorana fermion is a hypothetical particle which is its own antiparticle. The search for Majorana fermions goes back to the early days of relativistic quantum mechanics.

\subsection{Their origin in particle physics}
\label{particlephysics}

The notion of an antiparticle originated with Paul Dirac's 1930 interpretation of the negative-energy solutions of his relativistic wave equation for spin-$\tfrac{1}{2}$ particles \cite{Dir30}. The positive-energy solutions describe electrons, and the negative-energy solutions correspond to particles with the same mass and spin but opposite charge. The electron and its antiparticle, the positron, are related by a symmetry operation which takes the complex conjugate of the wave function. Particle and antiparticle can annihilate, producing a pair of photons. While the photon (described by a real bosonic field) is its own antiparticle, Dirac fermions are described by complex fields with distinct particle and antiparticle. 

In a paper published in 1937, shortly before his disappearance, Ettore Majorana questioned the need to introduce a distinct antiparticle for each particle \cite{Maj37}. The complex Dirac equation can be separated into a pair of real wave equations, each of which describes a real fermionic field \cite{Edd28}. Majorana suggested that neutral particles might be represented by a single real field, and concluded that ``there is now no need to assume the existence of antineutrons or antineutrinos''.

We since know that the neutron and antineutron are distinct particles, but the neutrino and antineutrino could well be the same particle observed in different states of motion \cite{Wil09}. It remains to be seen whether or not the Majorana fermion will go the way of the magnetic monopole, as a mathematical possibility that is not realized by Nature in an elementary particle.

\subsection{Their emergence in superconductors}
\label{condensedmatteremergence}

In condensed matter we can build on what Nature offers, by constructing quasiparticle excitations with exotic properties out of simpler building blocks. This happened for magnetic monopoles and it may happen for Majorana fermions. The strategy to use midgap excitations of a chiral \textit{p}-wave superconductor goes back two decades \cite{Kop91,Vol99,Sen00,Rea00,Kit01,Mot01,Das06} (with even earlier traces in the particle physics literature \cite{Jac81}). Recent developments in topological states of matter have brought this program closer to realization \cite{Has10,Qi10}.  

The electron and hole excitations of the superconductor play the role of particle and antiparticle. Electrons (filled states at energy $E$ above the Fermi level) and holes (empty states at $-E$ below the Fermi level) have opposite charge, but the charge difference of $2e$ can be absorbed as a Cooper pair in the superconducting condensate. At the Fermi level ($E=0$, in the middle of the superconducting gap), the eigenstates are charge neutral superpositions of electrons and holes.

That the midgap excitations of a superconductor are Majorana fermions follows from electron-hole symmetry: The creation and annihilation operators $\gamma^{\dagger}(E)$, $\gamma(E)$ for an excitation at energy $E$ are related by
\begin{equation}
\gamma(E)=\gamma^{\dagger}(-E).\label{gammaErelation}
\end{equation}
At the Fermi level $\gamma(0)\equiv\gamma=\gamma^{\dagger}$, so particle and antiparticle coincide. The anticommutation relation for Majorana fermion operators has the unusual form
\begin{equation}
\gamma_{n}\gamma_{m}+\gamma_{m}\gamma_{n}=2\delta_{nm}.\label{anticommutator}
\end{equation}
The operators of two Majoranas anticommute, as for any pair of fermions, but the product $\gamma_{n}^{2}=1$ does not vanish.

Like in the particle physics context, these are just formal manipulations if the state is degenerate --- since a Dirac fermion operator $a=\frac{1}{2}(\gamma_{1}+i\gamma_{2})$ is fully equivalent to a pair of Majorana operators. Spin degeneracy, in particular, needs to be broken in order to realize an unpaired Majorana fermion. The early proposals \cite{Kop91,Vol99,Sen00,Rea00,Kit01,Mot01,Das06} were based on an unconventional form of superconductivity, in which only a single spin band is involved. Such spin-triplet, \textit{p}-wave pairing is fragile, easily destroyed by disorder. Much of the recent excitement followed after Liang Fu and Charles Kane showed that conventional spin-singlet, \textit{s}-wave superconductivity could be used, in combination with the strong spin-orbit coupling of a topological insulator \cite{Fu08}.

\begin{figure}[tb]
\includegraphics[width=0.8\linewidth]{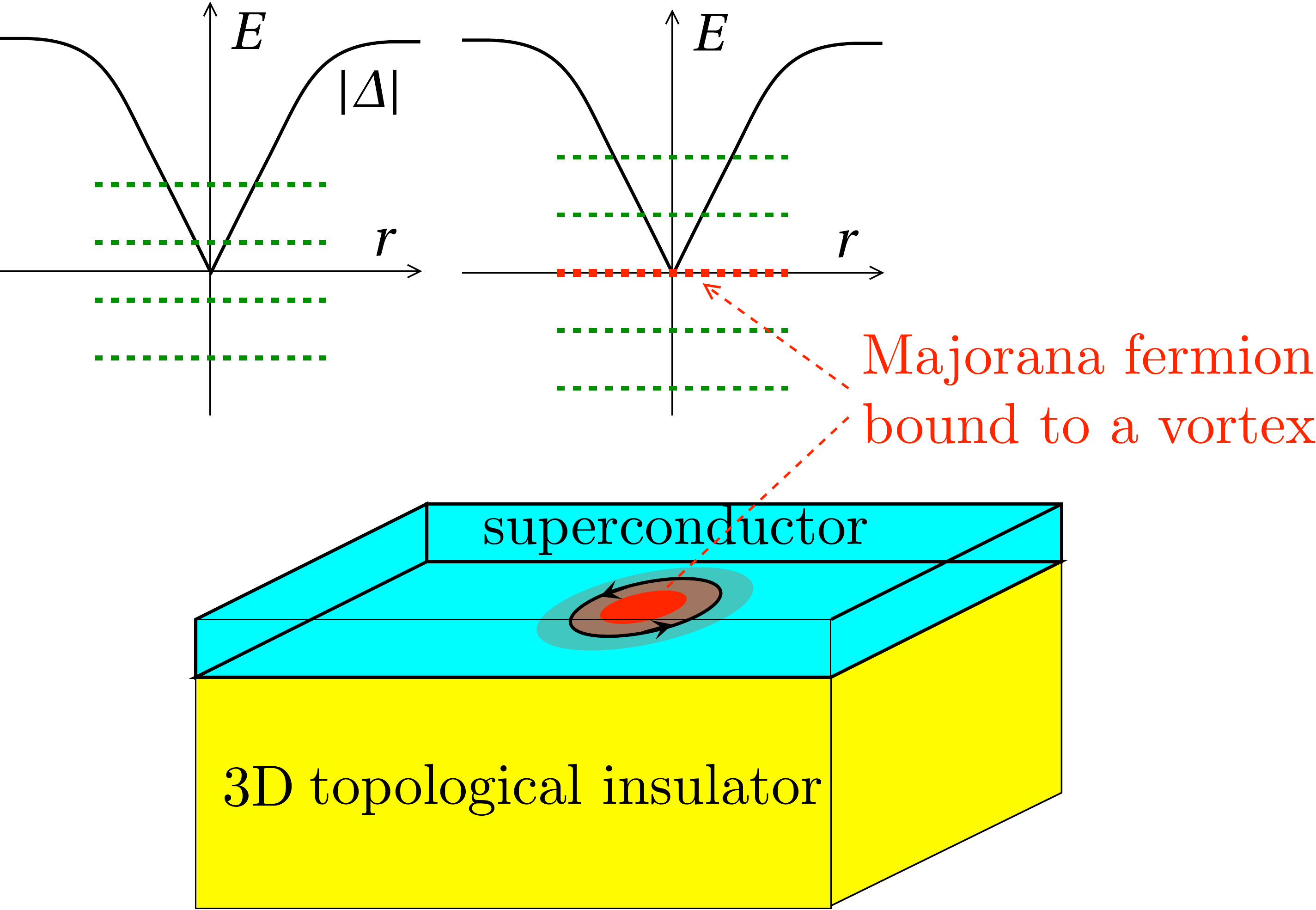}
\caption{\label{fig_vortices}
Profile of the superconducting pair potential $\Delta(r)$ in an Abrikosov vortex (solid curve) and bound electron-hole states in the vortex core (dashed lines). The left graph shows the usual sequence of levels in an \textit{s}-wave superconductor, arranged symmetrically around zero energy. The right graph shows the level sequence when superconductivity is induced on the surface of a 3D topological insulator, with a nondegenerate state at $E=0$. This midgap state is a Majorana fermion.
}
\end{figure}

The basic mechanism is illustrated in Fig.\ \ref{fig_vortices}. A three-dimensional (3D) topological insulator has an insulating bulk and a metallic surface \cite{Has10,Qi10}. The 2D surface electrons are massless Dirac fermions, very much like in graphene --- but without the spin and valley degeneracies of graphene. A superconductor deposited on the surface opens an excitation gap, which can be closed locally by a magnetic field. The magnetic field penetrates as an Abrikosov vortex, with subgap states $E_{n}=(n+\alpha)\delta$, $n=0,\pm 1,\pm 2,\ldots$, bound to the vortex core \cite{Car64}. (The level spacing $\delta\simeq \Delta^{2}/E_{F}$ is is determined by the superconducting gap $\Delta$ and the Fermi energy $E_{F}$.) Electron-hole symmetry restricts $\alpha$ to the values $0$ or $1/2$. For $\alpha=0$ the zero-mode $E_{0}=0$ would be a Majorana fermion in view of Eq.\ \eqref{gammaErelation}, but one would expect zero-point motion to enforce $\alpha=1/2$.

While $\alpha=1/2$ indeed holds for the usual massive electrons and holes, 2D massless Dirac fermions have $\alpha=0$ --- as discovered by Roman Jackiw and Paolo Rossi \cite{Jac81}. The reader familiar with graphene may recall the appearance of a Landau level at zero energy, signifying quantization of cyclotron motion without the usual $\frac{1}{2}\hbar\omega_{c}$ offset from zero-point motion \cite{McC56}. The absence of a $\frac{1}{2}\delta$ offset in an Abrikosov vortex has the same origin. Massless Dirac fermions have their spin pointing in the direction of motion. A closed orbit produces a phase shift of $\pi$ from the $360^{\circ}$ rotation of the spin. This Berry phase adds to the phase shift of $\pi$ in the Bohr-Sommerfeld quantization rule, converting destructive interference at $E=0$ into constructive interference and shifting the offset $\alpha$ from $1/2$ to $0$.

\subsection{Their potential for quantum computing}
\label{quantumcomputing}

\begin{figure}[tb]
\includegraphics[width=0.7\linewidth]{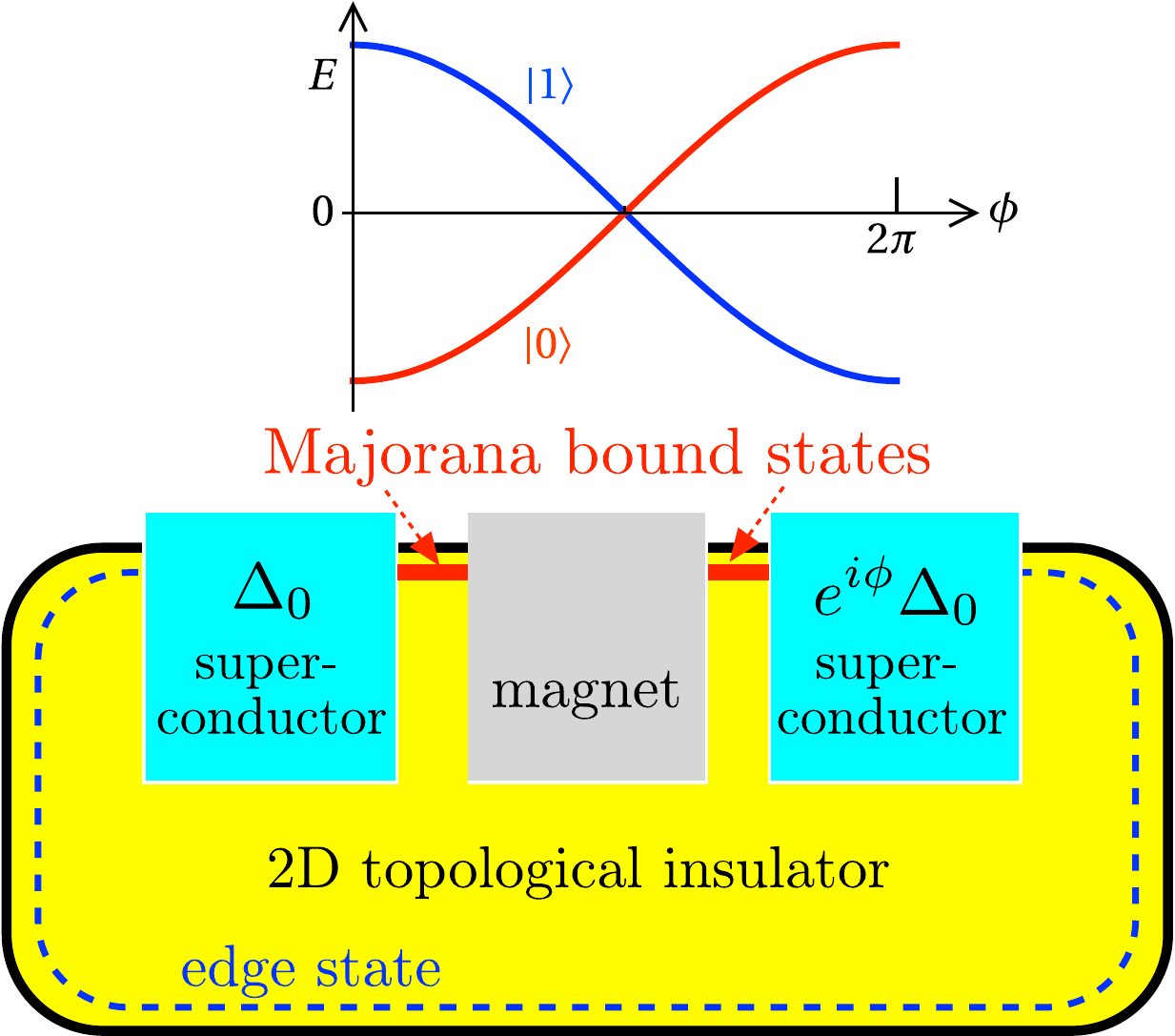}
\caption{\label{fig_QSHedge}
Top view of a 2D topological insulator, contacted at the edge by two superconducting electrodes separated by a magnetic tunnel junction. A pair of Majorana fermions is bound by the superconducting and magnetic gaps. The tunnel splitting of the bound states depends $\propto\cos(\phi/2)$ on the superconducting phase difference $\phi$, as indicated in the plot. The crossing of the levels at $\phi=\pi$ is protected by quasiparticle parity conservation.
}
\end{figure}

The idea to store quantum information in Majorana fermions originates from Alexei Kitaev \cite{Kit01}. We illustrate the basic idea in Fig.\ \ref{fig_QSHedge} in the context of a 2D topological insulator \cite{Fu09,Nil08}, one dimension lower than in Fig.\ \ref{fig_vortices}. The massless Dirac fermions now propagate along a 1D edge state, again with the spin pointing in the direction of motion. (This is the helical edge state responsible for the quantum spin Hall effect.) A Majorana fermion appears as a zero-mode at the interface between a superconductor (S) and a magnetic insulator (I).

Fig.\ \ref{fig_QSHedge} shows two zero-modes coupled by tunneling in an SIS junction, forming a two-level system (a qubit). The two states $|1\rangle$ and $|0\rangle$ of the qubit are distinguished by the presence or absence of an unpaired quasiparticle. For well-separated Majoranas, with an exponentially small tunnel splitting, this is a \textit{nonlocal} encoding of quantum information: Each zero-mode by itself contains no information on the quasiparticle parity.

Dephasing of the qubit is avoided by hiding the phase in much the same way that one would hide the phase of a complex number by separately storing the real and imaginary parts. The complex Dirac fermion operator $a=\frac{1}{2}(\gamma_{1}+i\gamma_{2})$ of the qubit is split into two real Majorana fermion operators $\gamma_{1}$ and $\gamma_{2}$. The quasiparticle parity $a^{\dagger}a=\frac{1}{2}(1+i\gamma_{1}\gamma_{2})$ is only accessible by a joint measurement on the two Majoranas. 

While two Majoranas encode one qubit, $2n$ Majoranas encode the quantum information of $n$ qubits in $2^{n}$ nearly degenerate states. Without these degeneracies, the adiabatic evolution of a state $\Psi$ along a closed loop in parameter space would simply amount to multiplication by a phase factor, $\Psi\mapsto e^{i\alpha}\Psi$, but now the operation may result in multiplication by a unitary matrix, $\Psi\mapsto U\Psi$. Because matrix multiplications do not commute, the order of the operations matters. This produces the non-Abelian statistics discovered by Gregory Moore and Nicholas Read \cite{Moo91}, in the context of the fractional quantum Hall effect, and by Read and Dmitry Green \cite{Rea00}, in the context of \textit{p}-wave superconductors.

The adiabatic interchange (braiding) of two Majorana bound states is a non-Abelian unitary transformation of the form
\begin{equation}
\Psi\mapsto \exp\left(i\frac{\pi}{4}\sigma_{z}\right)\Psi,\label{braiding}
\end{equation}
with $\sigma_{z}$ a Pauli matrix acting on the qubit formed by the two interchanging Majoranas \cite{Nay96,Iva01}. Two interchanges return the Majoranas to their starting position, but the final state $i\sigma_{z}\Psi$ is in general not equivalent to the initial state $\Psi$.

An operation of the form \eqref{braiding} is called \textit{topological}, because it is fully determined by the topology of the braiding; in particular, the coefficient in the exponent is precisely $\pi/4$. This could be useful for a quantum computer, even though not all unitary operations can be performed by the braiding of Majoranas \cite{Kit03,Nay08}.

Before closing this section, we emphasise that the object exhibiting non-Abelian statistics is not the Majorana fermion by itself, but the Majorana fermion bound to a topological defect (a vortex in Fig.\ \ref{fig_vortices}, the SI interface in Fig.\ \ref{fig_QSHedge}, or an $e/4$ quasiparticle in the fractional quantum Hall effect). The combined object is referred to as an Ising anyon in the literature on topological quantum computation \cite{Nay08}. A free Majorana fermion (such as may be discovered in particle physics) has ordinary fermionic statistics --- it is not an Ising anyon. The same applies to unbounded Majorana fermions at the edge or on the surface of a topological superconductor. 

In what follows we will concentrate on the Majoranas bound to a topological defect, because of their exotic statistics. We could have included the Ising anyons in the fractional quantum Hall effect, but because that topic is already very well reviewed \cite{Ste08}, we limit ourselves to the superconducting implementations.

\section{How to make them}
\label{howtomakethem}

The route to Majorana fermions in superconductors can follow a great variety of pathways. The growing list of proposals includes Refs.\  
\onlinecite{Kop91,Vol99,Sen00,Rea00,Kit01,Mot01,Das06,Fu08,Sen01,Fu09,Nil08,Sat09b,Lin10,Sau10,Ali10,Lee09,Lut10,Ore10,Neu10,Sat10b,Sat10,Qi10b,Duc11,Ser11,Wen11,Chu11,Mao11a,Gan11,Hos11,Bla11,Coo11,Mao11b,Tsv11,Jia12,Den11,Cho11,Mar11,Kja11,Sau11b,Sau11c,Sau11d,Won11,Kli12,Pot12,Tak12}. There are so many ways to make Majorana fermions because the requirements are so generic: Take a superconductor, remove degeneracies by breaking spin-rotation and time-reversal symmetries, and then close and reopen the excitation gap. As the gap goes through zero, Majorana fermions emerge as zero-modes bound to magnetic or electrostatic defects \cite{Vol03,Teo10}. We summarize the main pathways, and refer to a recent review \cite{Ali12} for a more detailed discussion.

\subsection{Shockley mechanism}
\label{Shockleymech}

\begin{figure}[tb]
\includegraphics[width=0.7\linewidth]{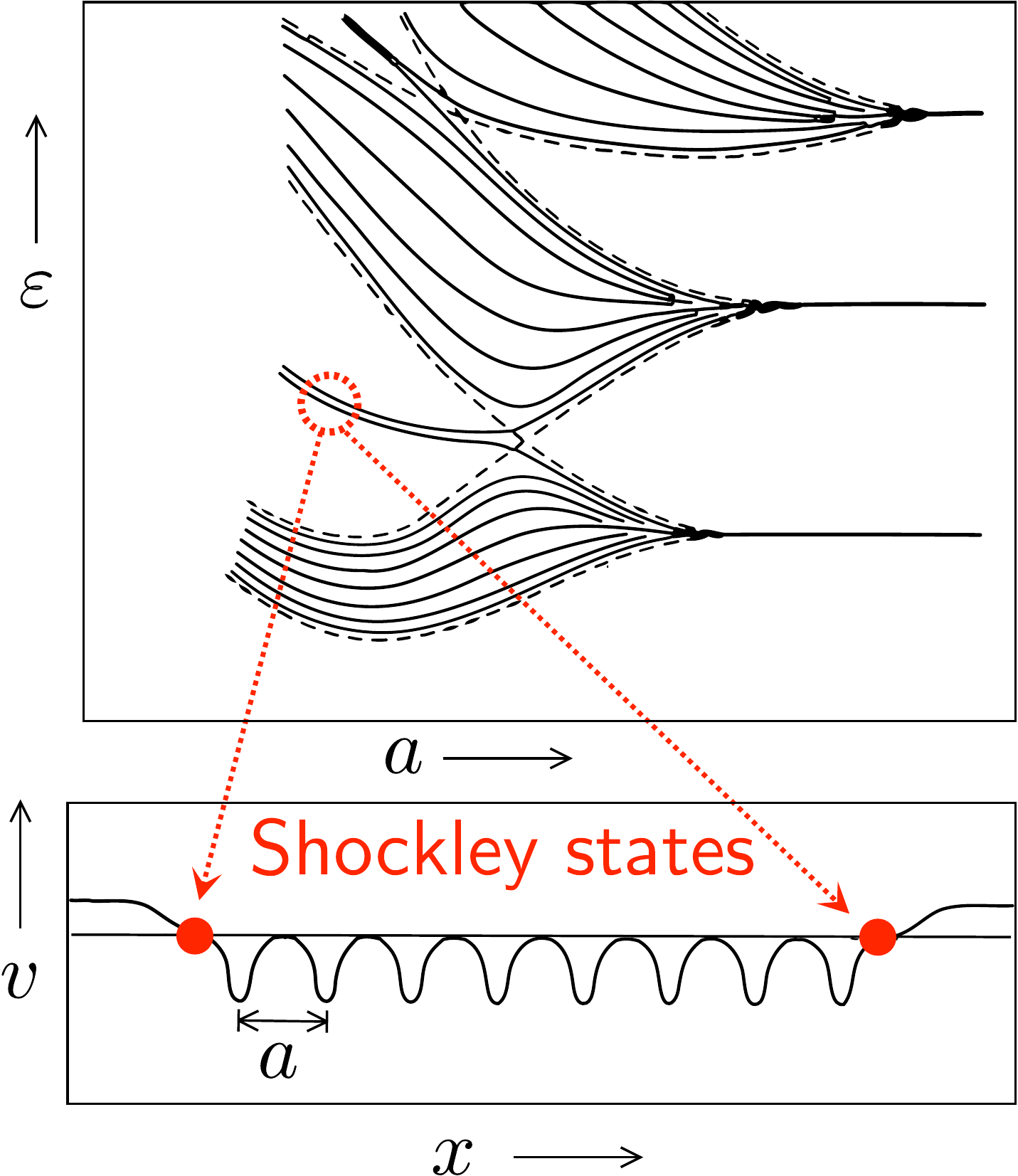}
\caption{\label{fig_Shockley}
Illustration of the Shockley mechanism for the formation of bound states at the end points of an atomic chain. The lower panel shows the potential profile along the chain and the upper panel shows the corresponding energy levels as a function of the atomic separation $a$. The end states appear upon the closing and reopening of the band gap. Figure adapted from Ref.\ \onlinecite{Sho39}.
}
\end{figure}

\begin{figure}[tb]
\includegraphics[width=1.0\linewidth]{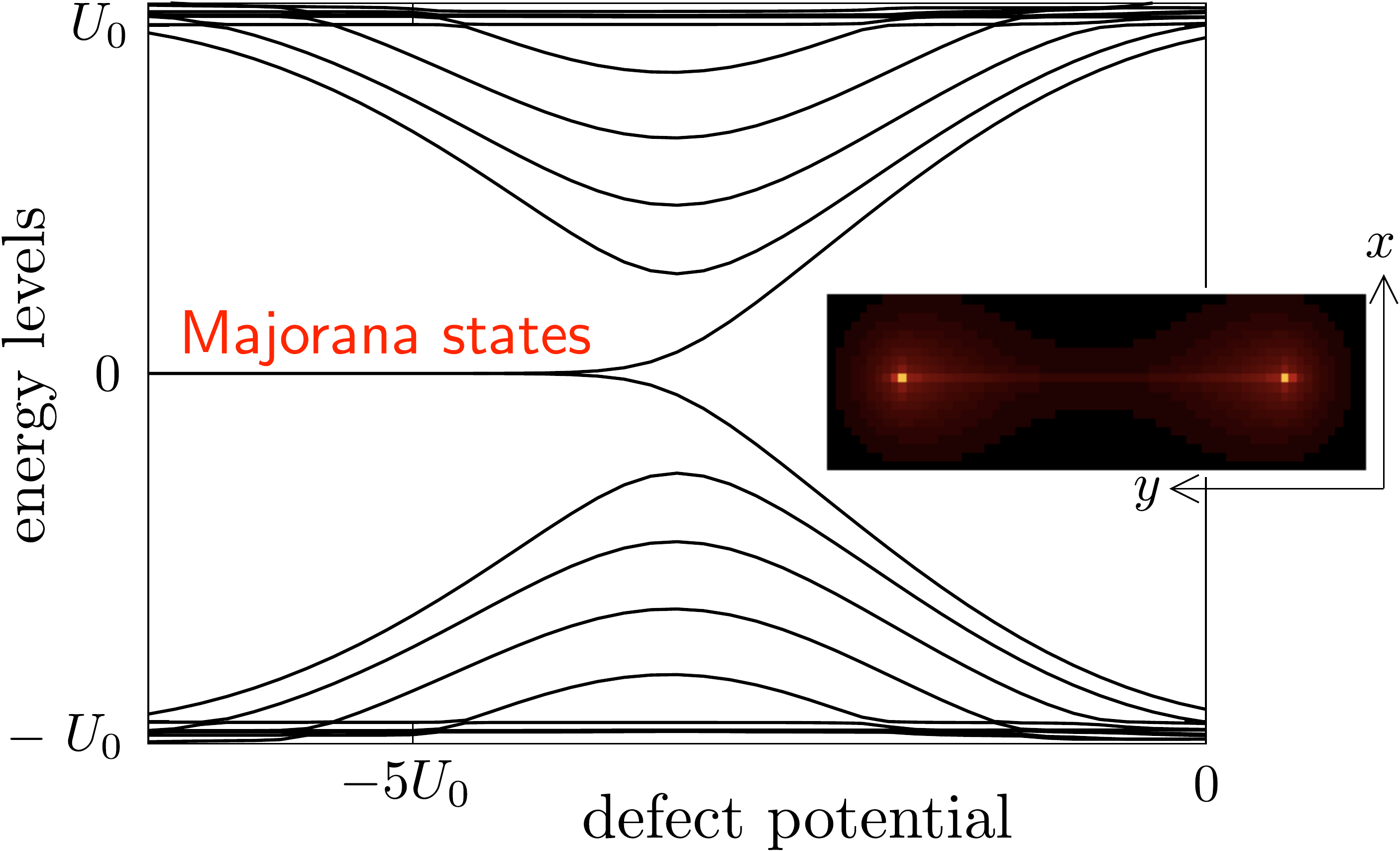}
\caption{\label{fig_MS}
Emergence of a pair of zero-energy Majorana states in a model calculation of a chiral \textit{p}-wave superconductor containing an electrostatic line defect. The gap closes and reopens as the defect potential $U_{0}+\delta U$ is made more and more negative, at fixed positive background potential $U_0$. The inset shows the probability density of the zero-mode in the 2D plane of the superconductor, with the line defect along the $y$-axis. Figure adapted from Ref.\ \onlinecite{Wim10}.
}
\end{figure}

From this general perspective, Majorana bound states can be understood as superconducting counterparts of the Shockley states from surface physics \cite{Sho39,Wim10}. The closing and reopening of a band gap in a chain of atoms leaves behind a pair of states in the gap, bound to the end points of the chain, see Fig.\ \ref{fig_Shockley}. Shockley states are unprotected and can be pushed out of the band gap by local perturbations. In a superconductor, in contrast, particle-hole symmetry requires the spectrum to be $\pm E$ symmetric, so an isolated bound state is constrained to lie at $E = 0$ and cannot be removed by any local perturbation, see Fig.\ \ref{fig_MS}.

The closing of the excitation gap, followed by its reopening with opposite sign, is a topological phase transition. The phase transition is called topological, in distinction to thermodynamic, because the sign ${\cal Q}=\pm 1$ of the gap cannot be seen in thermodynamic properties. This so-called $\mathbb{Z}_{2}$ topological quantum number counts the parity of the number of Majorana fermions bound to the defect. Only an odd number of Majoranas (${\cal Q}=-1$) produces a stable zero-mode. A defect with ${\cal Q}=-1$ is called topologically nontrivial, while for ${\cal Q}=+1$ it is called topologically trivial.

Because the Majorana fermions are constructed from ordinary Dirac electrons, an unpaired Majorana at a topologically nontrivial defect must have a counterpart somewhere else in the system. The two Majoranas are evident in Figs.\ \ref{fig_QSHedge} and \ref{fig_MS}. In Fig.\ \ref{fig_vortices} the second Majorana extends along the outer perimeter of the superconductor. One could try to eliminate this second Majorana by covering the entire topological insulator by a superconductor. But then the flux line would intersect the superconductor at two points, producing again a pair of Majoranas.

Let us see how these topological phase transitions appear in some representative systems.

\subsection{Chiral \textit{p}-wave superconductors}
\label{chiralpwave}

The closing and reopening of the gap in Fig.\ \ref{fig_MS} is described by the Bogoliubov-De Gennes Hamiltonian of a 2D chiral \textit{p}-wave superconductor,
\begin{equation}
H=\begin{pmatrix}
U+p^{2}/2m&\Delta (p_x-ip_y)\\
\Delta^{\ast}(p_x+ip_y)&-U-p^{2}/2m
\end{pmatrix}.\label{Hpwave}
\end{equation}
The diagonal elements give the electrostatic energy $\pm U(\bm{r})$ and kinetic energy $\pm p^2/2m$ of electrons and holes. (Energies are measured relative to the Fermi level.) The off-diagonal elements couple electrons and holes via the superconducting pair potential, which has the chiral \textit{p}-wave orbital symmetry $\propto p_x\pm ip_y$. (For equal-spin triplet pairing the spin degree of freedom can be omitted.)

\begin{figure}[tb]
\includegraphics[width=1\linewidth]{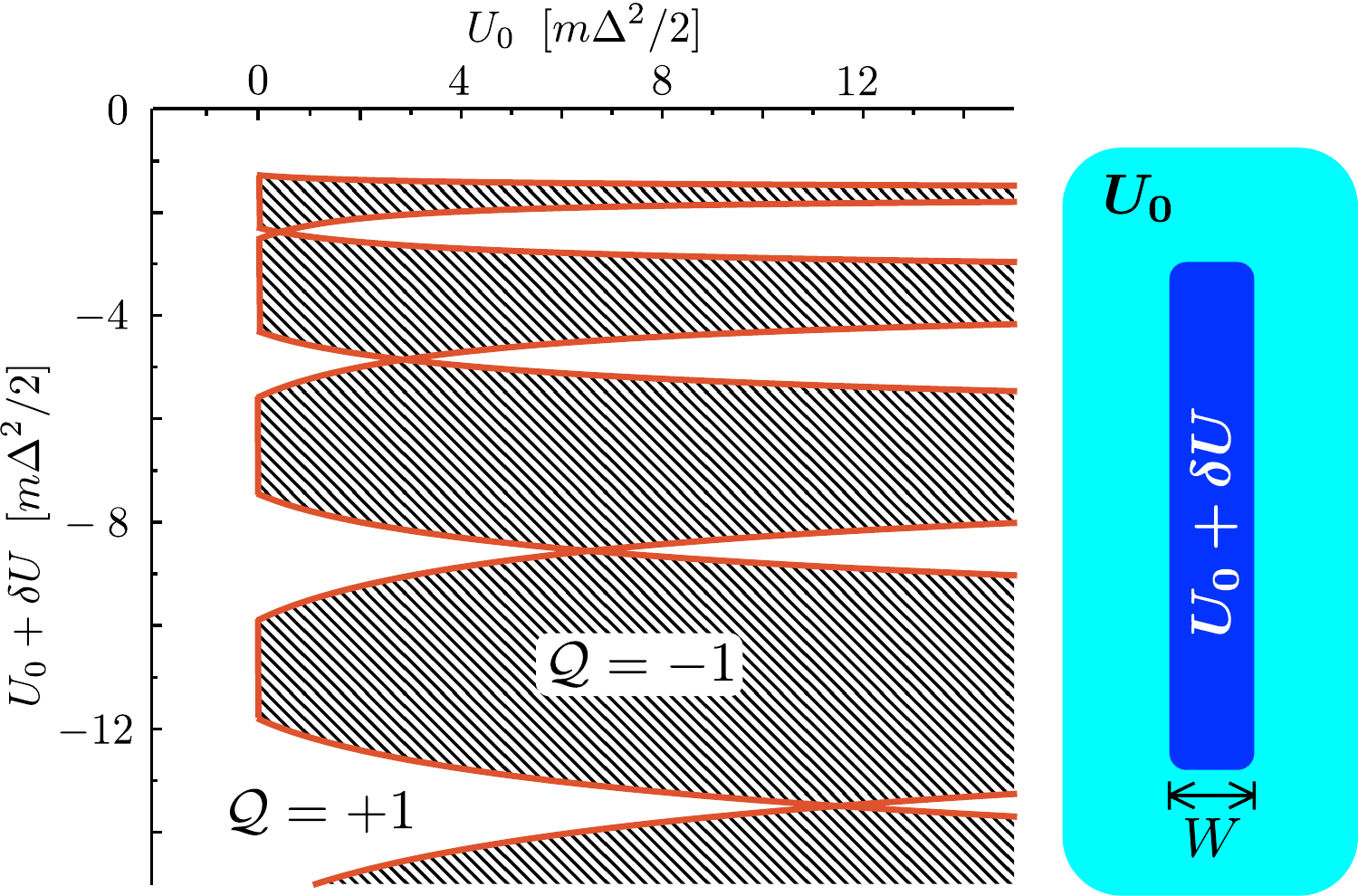}
\caption{\label{fig_pwavestrip}
The red solid curves locate the closing of the excitation gap of an electrostatic line defect (width $W=4\hbar/m\Delta$) in a chiral \textit{p}-wave superconductor, described by the Hamiltonian \eqref{Hpwave}. In the shaded regions the defect is topologically nontrivial (${\cal Q}=-1$), with Majorana states bound to the two ends. Figure adapted from Ref.\ \onlinecite{Wim10}.
}
\end{figure}

The line defect is constructed by changing $U$ from the background value $U_0$ to $U_{0}+\delta U$ in a strip of width $W$. As $\delta U$ is varied, multiple closings and reopenings of the gap appear, see Fig.\ \ref{fig_pwavestrip}. This is the 2D generalization \cite{Wim10,Pot10,Lut11,Zho11} of the 1D Kitaev chain \cite{Kit01}. The gap closing is a result of destructive interference of transverse modes in the strip. Each new mode is associated with one closing-reopening of the gap, so that the defect is topologically nontrivial  (${\cal Q}=-1$) for an odd number of modes and trivial  (${\cal Q}=+1$) for an even number of modes.

Electrostatic line defects are one way of producing Majorana fermions in a chiral \textit{p}-wave superconductor. Magnetic vortices are another way \cite{Vol99,Rea00,Sto06,Kra09,Mol11,Rak11}, those defects are topologically nontrivial for $U<0$. Strontium ruthenate ($\text{Sr}_2\text{RuO}_4$) is a candidate \textit{p}-wave material to observe the predicted zero modes \cite{Das06}.

\subsection{Topological insulators}
\label{topins}

In a topological insulator the closing and reopening of the band gap is a consequence of strong spin-orbit coupling, which inverts the order of conduction and valence bands \cite{Has10,Qi10}. The surface of a topological insulator supports nondegenerate, massless Dirac fermions, with Hamiltonian
\begin{equation}
H_{0}=v_{F}\bm{p}\cdot\bm{\sigma}+U+M\sigma_{z}.\label{HD}
\end{equation}
A 3D topological insulator, such as $\text{Bi}_{2}\text{Se}_{3}$ or $\text{Bi}_{2}\text{Te}_{3}$, has 2D Dirac fermions on the surface, while a 2D topological insulator, such as a HgTe/CdTe or InAs/GaSb quantum well, has 1D Dirac fermions along the edge. The term $\bm{p}\cdot\bm{\sigma}$ represents $p_{x}\sigma_{x}+p_{y}\sigma_{y}$ or $p_{x}\sigma_{x}$ for surface or edge Dirac fermions, respectively. The extra term $M\sigma_{z}$ accounts for the exchange energy from a magnetic insulator.

As illustrated in Figs.\ \ref{fig_vortices} and \ref{fig_QSHedge}, both the surface and edge Dirac fermions can give rise to Majorana bound states \cite{Fu08}. The Bogoliubov-De Gennes Hamiltonian that describes these zero modes has the form
\begin{equation}
H=\begin{pmatrix}
H_{0}&\Delta\\
\Delta^{\ast}&-\sigma_{y}H^{\ast}_{0}\sigma_{y}
\end{pmatrix}.\label{HBdG}
\end{equation}
The diagonal contains the electron and hole Dirac Hamiltonians \eqref{HD}, related to each other by the time-reversal operation $H_{0}\mapsto\sigma_{y}H^{\ast}_{0}\sigma_{y}$ (which inverts $\bm{p}$ and $\bm{\sigma}$). The pair potential $\Delta$ (which can be complex as a result of a magnetic vortex) is induced by \textit{s}-wave superconductivity, so it is momentum independent.

\begin{figure}[tb]
\includegraphics[width=0.8\linewidth]{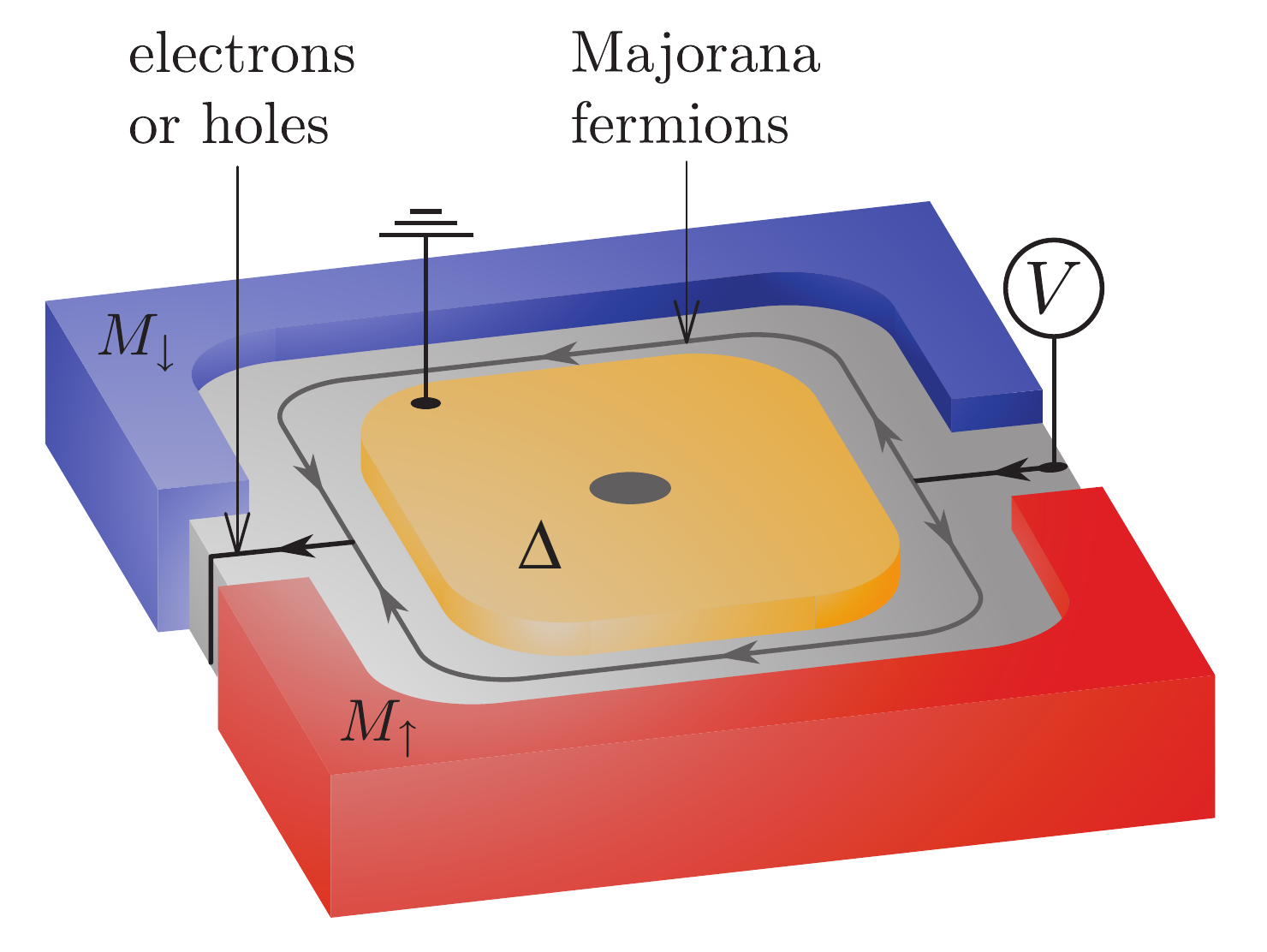}
\caption{\label{fig_DM}
Dirac-to-Majorana fermion converter on the surface of a 3D topological insulator. Arrows indicate the propating modes at the interface between a superconductor and a magnetic insulator and at the magnet-magnet interface. An electron (Dirac fermion) injected by the voltage source at the right is split into a pair of Majorana fermions. These fuse at the left, either back into an electron or into a hole, depending on whether the superconductor contains an even or an odd number of vortices. The recombination as a hole adds a Cooper pair to the superconductor, which can be detected in the current to ground. Figure adapted from Ref.\ \cite{Akh09}.
}
\end{figure}

Fig. \ref{fig_DM} illustrates the conversion of Dirac fermions into Majorana fermions on the surface of a 3D topological insulator \cite{Fu09b,Akh09}. Part of the surface is covered with a superconductor ($M=0,\Delta\neq 0$) and part is covered with a magnetic insulator ($\Delta=0,M\neq 0$). Both $\Delta$ and $M$ open a gap for the surface states. The gap closes at the $M$--$\Delta$ interface and also at the $M_\uparrow$--$M_\downarrow$ interface between opposite magnetic polarizations (from $M_\uparrow>0$ to $M_\downarrow<0$). The 1D interface states propagate only in a single direction, much like the chiral edge states of the quantum Hall effect.

The $M_\uparrow$--$M_\downarrow$ interface leaves electrons and holes uncoupled, so there are two modes at the Fermi level, one containing electrons and one containing holes. These are Dirac fermions, with distinct creation and annihilation operators $a^{\dagger}$ and $a$. The electron-hole degeneracy is broken at the $M$--$\Delta$ interface, which supports only a single mode $\gamma=\gamma^{\dagger}$ at the Fermi level. The $M_\uparrow$--$M_\downarrow$--$\Delta$ tri-junction splits an electron or hole into two Majorana fermions,
\begin{equation}
a\rightarrow \tfrac{1}{2}(\gamma_{1}+i\gamma_{2}),\;\;a^{\dagger}\rightarrow\tfrac{1}{2}(\gamma_{1}-i\gamma_{2}).\label{agammasplitting}
\end{equation}

The inverse process, the fusion of two Majorana fermions into an electron or hole, happens after the Majorana fermions have encirled the superconductor and picked up a relative phase shift. At the Fermi level this phase shift is entirely determined by the parity of the number $n$ of vortices in the superconductor. For even $n$ the fusion conserves the charge, while for odd $n$ a charge $2e$ is added as a Cooper pair to the superconductor (so an electron is converted into a hole and \textit{vice versa}). The two processes can be distinguished experimentally by measuring the current to ground in the superconductor. Since each vortex binds one Majorana fermion (see Fig.\ \ref{fig_vortices}), the structure of Fig.\ \ref{fig_DM} can be seen as an electrical interferometer in which mobile Majoranas measure the parity of the number of Majorana bound states.

\subsection{Semiconductor heterostructures}
\label{semincondhetero}

The gap inversion in a topological insulator happens without superconductivity. One might alternatively try to directly invert the superconducting gap. In an \textit{s}-wave superconductor a magnetic field closes the gap, but how to reopen it? A promising strategy is to rely on the competing effects of a spin-polarizing Zeeman energy and a depolarizing spin-orbit coupling \cite{Sau10,Ali10,Sat09}. 

\begin{figure*}[tb]
\includegraphics[width=0.7\linewidth]{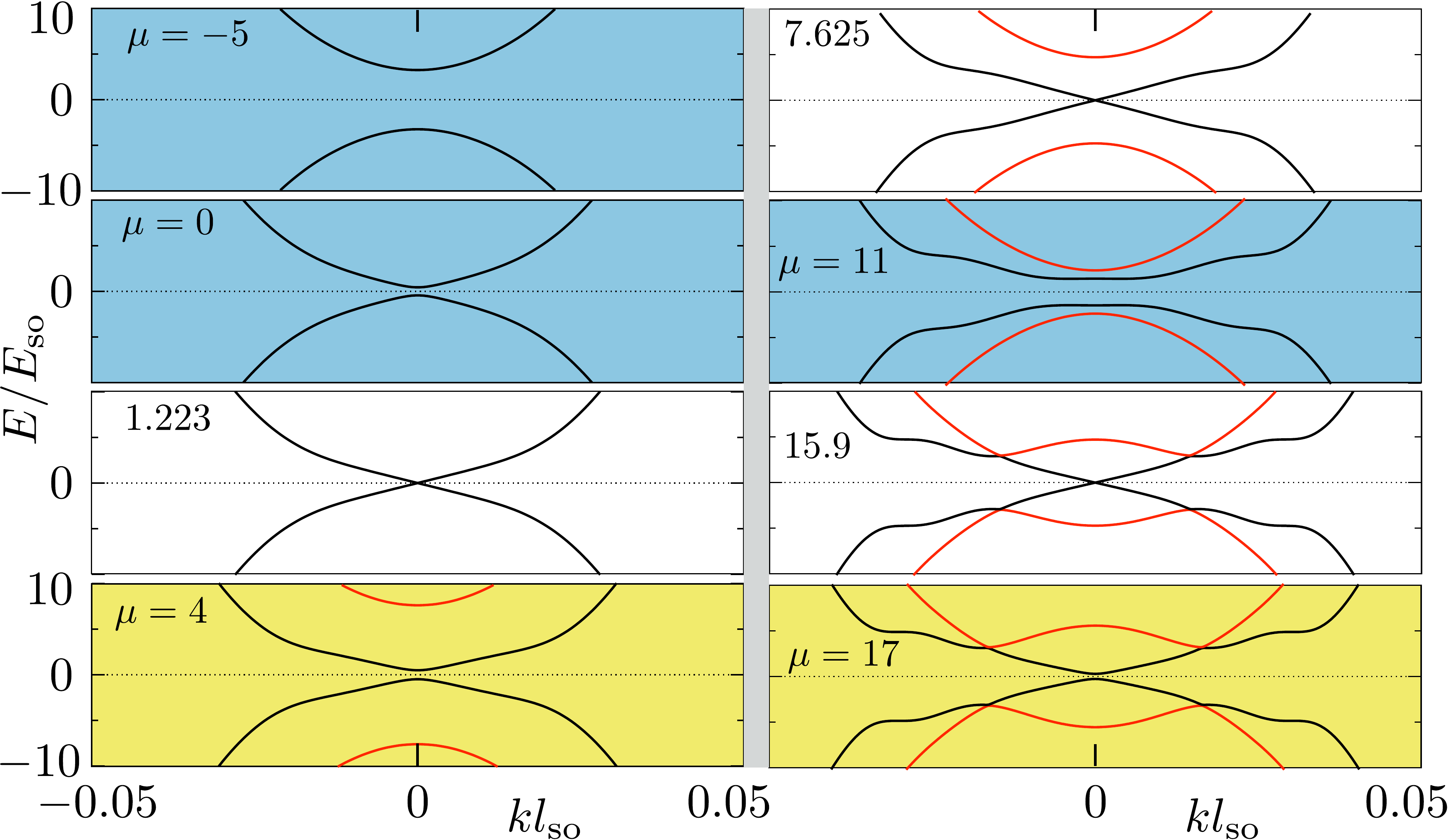}
\caption{\label{fig_dispersion}
Closing and reopening of the band gap in the Hamiltonian \eqref{HRdef} of a 2D semiconducting nanowire (width $W=l_{\rm so}$) on a superconducting substrate in a parallel magnetic field ($\Delta=10\,E_{\rm so}$, $E_{Z}=10.5\,E_{\rm so}$). The eight panels show the excitation energy near the Fermi level ($E=0$) as a function of the wave vector $k$ along the nanowire, for different values of the chemical potential $\mu\equiv -U$ (listed in units of the spin-orbit coupling energ $E_{\rm so}$). The colors blue or yellow of the panels indicate that the system is in a topologically trivial or nontrivial phase, respectively. The topological phase transition occurs in the uncolored panels. The nanowire supports Majorana bound states in the yellow panels. Data supplied by M. Wimmer.
}
\end{figure*}

The 2D electron gas of a semiconductor heterostructure, such as an InAs quantum well,  has a strong spin-orbit coupling from the Rashba effect. The orbital effect of a parallel magnetic field is suppressed, leaving only the spin-polarizing Zeeman effect. These two effects compete in the Hamiltonian
\begin{equation}
H_{0}=\frac{\bm{p}^{2}}{2m}+U+\frac{\alpha_{\rm so}}{\hbar}(\sigma_x p_y-\sigma_y p_x)+\tfrac{1}{2}g_{\rm eff}\mu_{B}B\sigma_{x}.\label{HRdef}
\end{equation}
Characteristic length and energy scales are $l_{\rm so} = \hbar^{2}/m\alpha_{\rm so}$ and $E_{\rm so} = m\alpha^{2}_{\rm so}/\hbar^{2}$. Typical values in InAs are $l_{\rm so} = 100\,{\rm nm}$, $E_{\rm so} = 0.1\,{\rm meV}$. The Zeeman energy is $E_{Z}=\frac{1}{2}g_{\rm eff}\mu_{B}B=1\,{\rm meV}$ at a magnetic field $B=1\,{\rm T}$. A superconducting proximity effect with a type-II superconductor like Nb is quite possible at these field strengths. The pair potential $\Delta$ induced in the 2D electron gas then couples electrons and holes via the Bogoliubov-De Gennes Hamiltonian \eqref{HBdG} [now with $H_{0}$ given by Eq.\ \eqref{HRdef}]. 

As discovered in Refs.\ \onlinecite{Lut10,Ore10}, the resulting band gap in a nanowire geometry closes and reopens upon variation of electron density (through a variation of $U$) or magnetic field (see Fig.\ \ref{fig_dispersion}). Majorana bound states at the two ends of the nanowire alternatingly appear and disappear at each of these topological phase transitions.

\section{How to detect them}
\label{howtodetectthem}

Majorana fermions modify the transport properties and the thermodynamic properties of the superconductor, providing ways to detect them. We summarize some of the signature effects of Majorana fermions, others can be found in Refs.\ \onlinecite{Nil08,Fu09b,Akh09,Tan09,Ben10,Asa10,Gro11,Akh11,Str11,Chu11b,Zit11,Mes11,Liu11,Wal11,Zoc12}.

\subsection{Half-integer conductance quantization}
\label{condquant}

\begin{figure}[tb]
\includegraphics[width=0.9\linewidth]{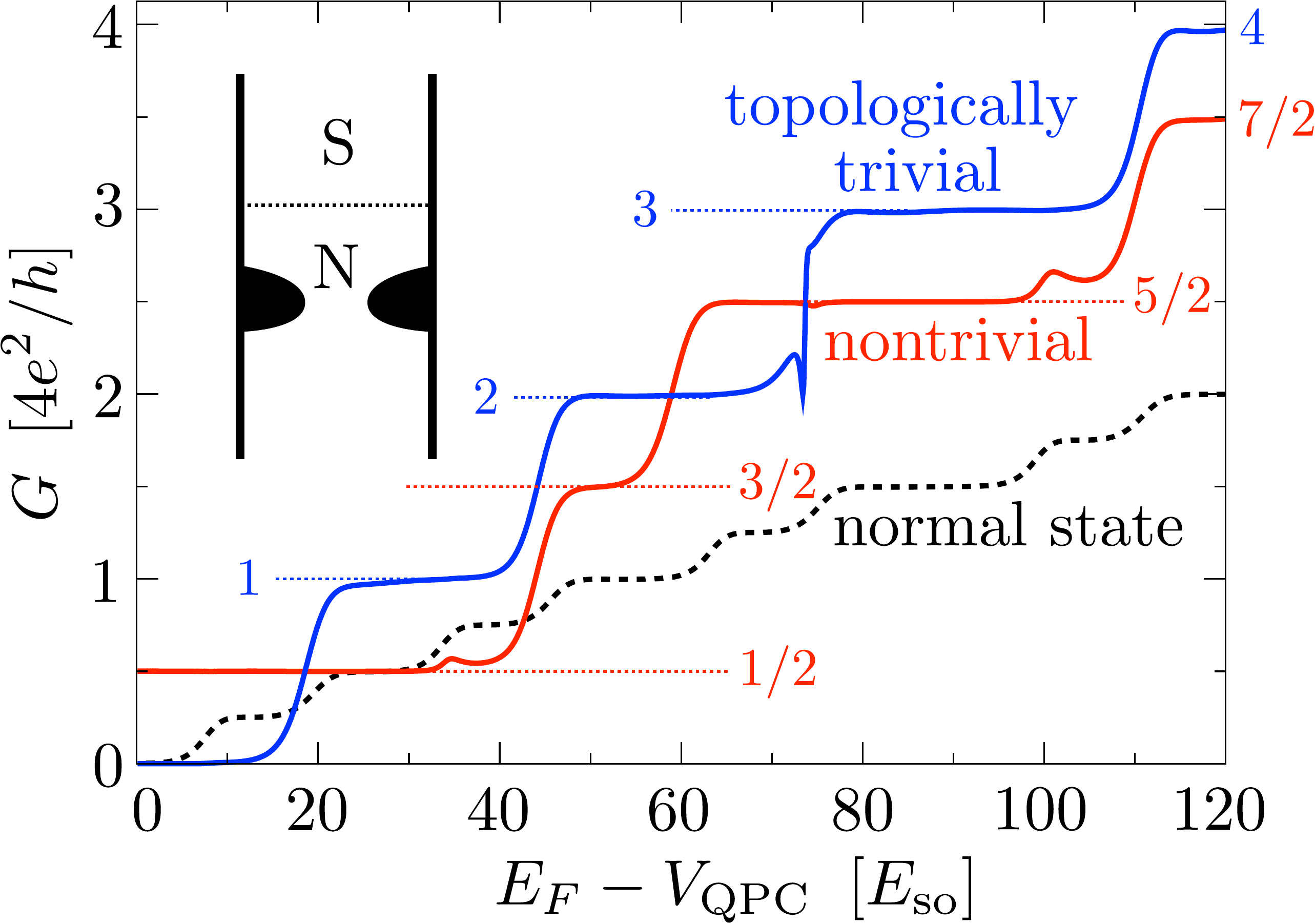}
\caption{\label{fig_QPC}
Solid curves: conductance of a ballistic NS junction, with the superconductor in a topologically trivial or non-trivial phase. The dotted curve is for an entirely normal system. The data is calculated from the model Hamiltonian \eqref{HRdef}. The point contact width is varied by varying the potential $V_{\rm QPC}$ inside the constriction at constant Fermi energy $E_{F}$. The dotted horizontal lines indicate the shift from integer to half-integer conductance plateaus upon transition from the topologically trivial to nontrivial phase. Figure adapted from Ref.\ \onlinecite{Wim11}.
}
\end{figure}

Tunneling spectroscopy is a direct method of detection of a Majorana bound state \cite{Bol07,Law09}: Resonant tunneling into the midgap state produces a conductance of $2e^{2}/h$, while without this state the conductance vanishes. A complication in the interpretation of tunneling spectroscopy is that the zero-bias peak may be obscured by resonances from subgap states at nonzero energy \cite{Fle10,Kel12}. A ballistic point contact provides a more distinctive signature of the topologically nontrivial phase \cite{Wim11}, through the half-integer conductance plateaus shown in Fig.\ \ref{fig_QPC}.

Both the tunneling and ballistic conductances can be understood from the general relation \cite{Blo82} between the conductance $G$ of a normal-metal--superconductor (NS) junction and the Andreev reflection eigenvalues $R_{n}$,
\begin{equation}
G=\frac{2e^{2}}{h}\sum_{n}R_{n}.\label{BTK}
\end{equation}
The $R_{n}$'s represent the probability for Andreev reflection in the $n$-th eigenmode at the Fermi level. The factor of two is not due to spin (which is included in the sum over $n$), but due to the fact that Andreev reflection of an electron into a hole doubles the current. 

There is no time-reversal symmetry, so Kramers degeneracy does not apply. Still, particle-hole symmetry requires that any $R_{n}$ is twofold degenerate (B\'{e}ri degeneracy \cite{Ber09}) --- with two exceptions: $R_{n}=0$ and $R_{n}=1$ may be nondegenerate. The nondegenerate Andreev reflection eigenvalue from a Majorana bound state is pinned to unity, contributing to the conductance a quantized amount of $2e^{2}/h$. All other fully Andreev reflected modes are twofold degenerate and contribute $4e^{2}/h$. The resulting conductance plateaus therefore appear at integer or half-integer multiples of $4e^{2}/h$, depending on whether the superconductor is topologically trivial or not.

The plateaus at $(n+1/2)\times 4e^{2}/h$ are reminiscent of the quantum Hall plateaus in graphene, and both originate from a zero mode, but the sensitivity to disorder is entirely different. The topological quantum number ${\cal Q}\in\mathbb{Z}$ for the quantum Hall effect, while ${\cal Q}\in\mathbb{Z}_{2}$ for a topological superconductor. The corresponding topological protection against disorder extends to all plateaus for the quantum Hall effect, but only to the lowest $n=0$ plateau for the topological superconductor.

It may appear paradoxical \cite{Sem07,Tew08} to have an electrical current flowing through a single Majorana bound state, since one Majorana fermion operator $\gamma$ represents only half of an electronic state.  However, the Hermitian operator $i(a+a^{\dagger})\gamma$ is a local coupling of Dirac and Majorana operators \cite{Bol07}, so electrical conduction can be a fully local process --- involving only one of the two spatially separated Majorana fermions.

\subsection{Nonlocal tunneling}
\label{nonlocal}

Nonlocal conduction involving both Majorana fermions becomes possible if there is a coupling between them \cite{Nil08,Fu10,Law09,Bos11,Pik11,Zaz11,Wan11,Did12}. The coupling term has the generic form $iE_{M}\gamma_{1}\gamma_{2}$, with eigenvalues $\pm E_{M}$. The energy $E_{M}$ may be a tunnel coupling due to overlap of wave functions, in which case it decays exponentially $\propto e^{-d/\xi_{0}}$ with the ratio of the separation $d$ of the Majoranas and the superconducting coherence length $\xi_{0}$. If the superconductor is electrically isolated (not grounded) and of small capacitance $C$, then the charging energy $E_{M}\simeq e^{2}/C$ provides a Coulomb coupling even without overlap of wave functions. (Recall that the two states of a pair of Majoranas are distinguished by the presence or absence of an unpaired quasiparticle, see Section \ref{quantumcomputing}.)

\begin{figure}[tb]
\includegraphics[width=0.9\linewidth]{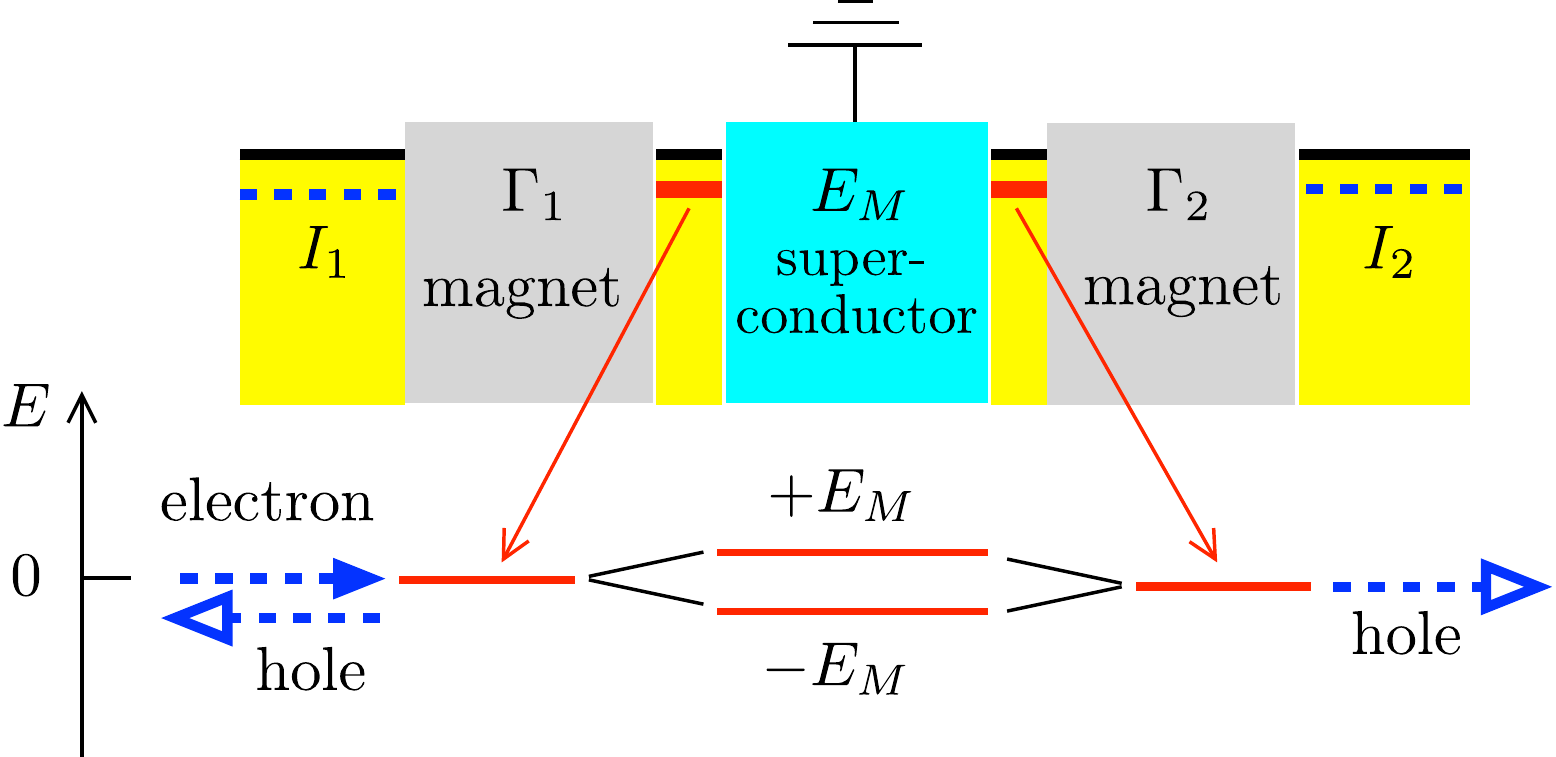}
\caption{\label{fig_nonlocal}
Majorana bound states (red) at the edge of a 2D topological insulator (cf.\ Fig.\ \ref{fig_QSHedge}), split into a pair of levels at $\pm E_{M}$ by a nonzero overlap. The levels are broadened due to a tunnel coupling $\Gamma_{1},\Gamma_{2}$ through the magnet to the outside edge state. An electron incident from the left on the grounded superconductor can be Andreev reflected as a hole, either locally (to the left) or nonlocally (to the right). Nonlocal Andreev reflection is equivalent to the splitting of a Cooper pair by the two Majoranas. For $\Gamma_{1},\Gamma_{2}\ll E_M$ local Andreev reflection is suppressed. Figure adapted from Ref.\ \onlinecite{Nil08}.
}
\end{figure}

Nonlocal tunneling processes appear if the level splitting $E_{M}$ is large compared to the level broadening $\Gamma_{1},\Gamma_{2}$. For a grounded superconductor the nonlocality takes the form of nonlocal Andreev reflection, which amounts to a splitting of a Cooper pair by the two Majorana bound states \cite{Nil08} (see Fig.\ \ref{fig_nonlocal}). The Cooper pair splitting can be detected in a noise measurement through a positive cross-correlation of the currents $I_{1}$ and $I_{2}$ to the left and right of the superconductor. 

For an electrically isolated superconducting island any charge transfer onto the island is forbidden by the charging energy, so there can be no Andreev reflection. An electron incident on one side of the island is either reflected to the same side or transmitted, still as an electron, to the other side. The nonlocality \cite{Fu10} now appears in the ratio of the reflection and transmission probabilities on resonance,
\begin{equation}
R/T=(\Gamma_{1}-\Gamma_{2})^{2}/(\Gamma_{1}+\Gamma_{2})^{2},\label{RTratio}
\end{equation}
which is independent of the size of the island. No matter how far the two Majoranas are separated, the charging energy couples them into a single electronic level. In particular, for identical tunnel couplings $\Gamma_{1}=\Gamma_{2}$ the electron is resonantly transmitted through the island with unit probability . 

\subsection{4$\bm{\pi}$-periodic Josephson effect}
\label{fourpi}

So far we discussed signatures of Majoranas in the electrical conduction out of equilibrium, in response to a voltage difference between the superconductor and a normal-metal electrode. In equilibrium an electrical current (supercurrent) can flow between two superconductors in the absence of any applied voltage. This familiar {\sc dc} Josephson effect \cite{Tin04} acquires a new twist \cite{Kit01,Fu09,Lut10,Ore10,Kwo03,Law11,Ios11,Bad11,Jia11,Hec11,San12,Pik12,Zaz12,Dom12} if the junction between the superconductors contains Majorana fermions, as in Fig. \ref{fig_QSHedge}.

Quite generally, the supercurrent $I_{J}$ is given by the derivative
\begin{equation}
I_{J}=\frac{2e}{\hbar}\frac{dE}{d\phi}\label{IJHJ} 
\end{equation}
of the energy $E$ of the Josephson junction with respect to the superconducting phase difference $\phi$. While in the conventional Josephson effect only Cooper pairs can tunnel (with probability $\tau\ll 1$), Majorana fermions enable the tunneling of single electrons (with a larger probability $\sqrt{\tau}$). The switch from $2e$ to $e$ as the unit of transferred charge between the superconductors amounts to a doubling of the fundamental periodicity of the Josephson energy, from $E\propto \cos\phi$ to $E\propto\cos(\phi/2)$. 

If the superconductors form a ring, enclosing a flux $\Phi$, the period of the flux dependence of the supercurrent $I_{J}$ doubles from $2\pi$ to $4\pi$ as a function of the Aharonov-Bohm phase $2e\Phi/\hbar$. This is the $4\pi$-periodic Josephson effect \cite{Kit01,Kwo03}. As a function of the enclosed flux, $I_{J}$ has the same $h/e$ periodicity as the persistent current $I_{N}$ through a normal-metal ring (radius $L$), but the size dependence is entirely different: While $I_{N}$ decays as $1/L$ or faster, $I_{J}$ has the $L$-independence of a supercurrent. 

Since the two branches of the $E$-$\phi$ relation differ by one unpaired quasiparticle (see Fig.\ \ref{fig_QSHedge}), external tunneling events which change the quasiparticle parity (so-called quasiparticle poisoning) restore the conventional $2\pi$-periodicity \cite{Fu09}. In a closed system, the $4\pi$-periodicity is thermodynamically stable, provided that the entire ring is in a topologically nontrivial state (to prevent quantum phase slips) \cite{Hec11}.

\subsection{Thermal metal-insulator transition}
\label{thermal}

Collective properties of Majorana fermions can be detected in the thermal conductance. Superconductors are thermal insulators, because the excitation gap $\Delta$ suppresses the energy transport by quasiparticle excitations at low temperatures $T_{0}\ll\Delta/k_{B}$. Disorder can create states in the gap, but these are typically localized. However, the Majorana midgap states in a topological superconductor can give rise to extended states, since they are all resonant at the Fermi level. This transforms a thermal insulator into a thermal metal \cite{Boc00,Cha02}.

The thermal metal-insulator transition is called a class D Anderson transition, in reference to a classification of disordered systems in terms of the presence or absence of time-reversal, spin-rotation, and particle-hole symmetry \cite{Alt97,Eve08,Ryu10}. Class D has only particle-hole symmetry. The chiral \textit{p}-wave superconductor is a two-dimensional system in class D. Its thermal transport properties in the absence of Majorana fermions are similar to the electrical transport properties in the quantum Hall effect (class A, all symmetries broken). The bulk is insulating while the boundary supports chiral (unidirectional) edge states that give rise to the thermal quantum Hall effect \cite{Sen00,Rea00}. The thermal analogue of the conductance quantum $e^2/h$ is $G_{0}=\pi^{2}k_{B}^{2}T_{0}/6h$.

\begin{figure}[tb]
\centerline{\includegraphics[width=1\linewidth]{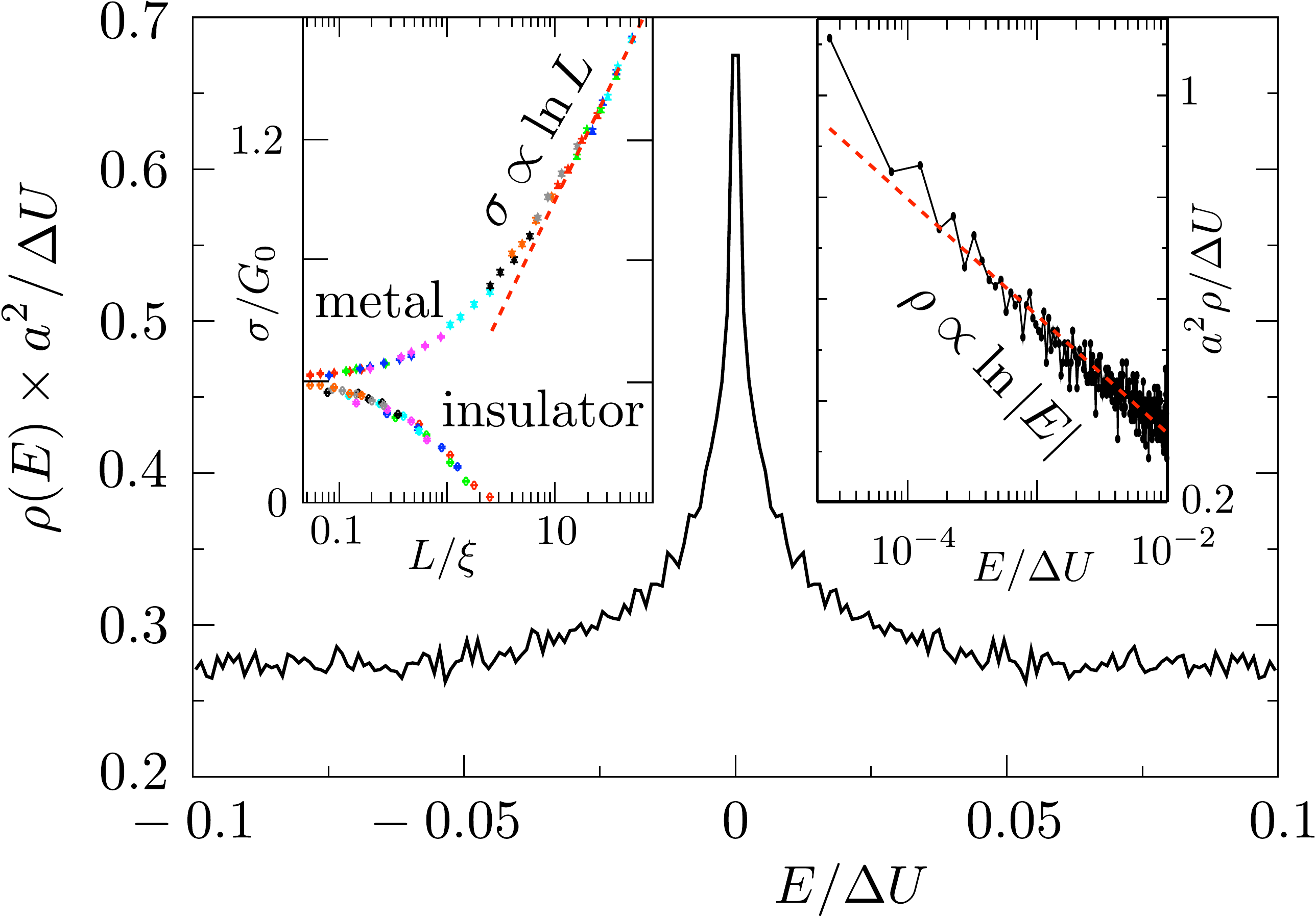}}
\caption{\label{fig_peak}
\textit{Main plot and right inset:} Average density of states $\rho$ in a model calculation of a chiral \textit{p}-wave superconductor with electrostatic disorder \cite{Wim10}. The Hamiltonian \eqref{Hpwave} is discretized on a lattice of size $400\,a\times 400\,a$ and the potential fluctuates randomly from site to site (r.m.s.\ $\Delta U$). Majorana fermions produce a midgap peak in the density of states. \textit{Left inset:} Average thermal conductivity $\sigma$ in a strip geometry of length $L$ and width $W=5L$, for the same Hamiltonian \eqref{Hpwave} but calculated with a different method of discretization \cite{Med10}. Data points of different color correspond to different disorder strengths $\Delta U$ and different scattering lengths $\xi$. Upon increasing disorder, a transition from insulating to metallic scaling is observed. In the metallic phase, the conductivity and density of states have a logarithmic dependence on, respectively, system size and energy.
}
\end{figure}

The correspondence between thermal and electrical quantum Hall effect breaks down in the presence of Majorana fermions. Their collective effect is illustrated in Fig.\ \ref{fig_peak}, obtained from the Hamiltonian \eqref{Hpwave} of a chiral \textit{p}-wave superconductor \cite{Wim10,Med10}. (Similar results have been obtained in other models of Majorana fermions \cite{Mil07,Lau11,Kra11}.) A randomly varying electrostatic potential creates a random arrangement of Majorana midgap states, via the Shockley mechanism of Fig.\ \ref{fig_MS}. The states are slightly displaced from $E=0$ by the overlap of wave functions. The resulting density of states has the logarithmic profile $\propto\ln |E|$, responsible for the logarithmic size dependence of the thermal conductivity \cite{Sen00},
\begin{equation}
\sigma=(G_{0}/\pi)\ln L+{\rm constant}.\label{sigmathermal}
\end{equation}
The thermal metal-insulator transition of Fig.\ \ref{fig_peak} has no electrical analogue.

\section{How to use them}
\label{computer}

Finding Majorana fermions in a superconductor is rewarding in and of itself. These particles might also provide a fundamentally new way to store and manipulate quantum information, with possible applications in a quantum computer.

\subsection{Topological qubits}
\label{topqubit}

In Section \ref{quantumcomputing} we considered a qubit formed out of a pair of Majorana fermions. The two states $|0\rangle$ and $|1\rangle$ of this elementary qubit differ by quasiparticle parity, which prevents the creation of a coherent superposition. For a quantum computation we combine two elementary qubits into a single logical qubit, consisting of four Majorana fermions \cite{Nay08}. Without loss of generality one can assume that the joint quasiparticle parity is even. The two states of the logical qubit are then encoded as $|00\rangle$ and $|11\rangle$. These two states have the same quasiparticle parity, so coherent superpositions are allowed.

An arbitrary state $|\Psi\rangle$ of the logical qubit has the form
\begin{equation}
|\Psi\rangle=\alpha|00\rangle+\beta|11\rangle,\;\;|\alpha|^{2}+|\beta|^{2}=1.\label{Psiqubit}
\end{equation}
Pauli matrices in the computational basis $|00\rangle$, $|11\rangle$ are bilinear combinations of the four Majorana operators,
\begin{equation}
\sigma_x=-i\gamma_{2}\gamma_{3}\;\;\sigma_{y}=i\gamma_{1}\gamma_{3},\;\;\sigma_{z}=-i\gamma_{1}\gamma_{2}.\label{Pauligamma}
\end{equation}
It is said that the qubit \eqref{Psiqubit} is topologically protected from decoherence by the environment \cite{Kit01}, because the bit-flip or phase-shift errors produced by the Pauli matrices \eqref{Pauligamma} can only appear if there is a coupling between pairs of Majorana fermions. The two types of coupling were discussed in Section \ref{nonlocal}: tunnel coupling when the Majoranas are separated by less than a coherence length, and Coulomb coupling when the Majoranas are on a superconducting island of small capacitance. 

The topological protection relies on the presence of a nonzero gap for quasiparticle excitations. Sub-gap excitations may exchange a quasiparticle with a Majorana fermion, provoking a bit-flip error. Error correction is possible if the sub-gap excitations remain bound to the Majorana fermion, in particular sub-gap excitations in a vortex core are not a source of decoherence \cite{Akh10,Gol11,Che11}. The topological protection does not apply if the superconductor is contacted by a gapless metal, allowing for the exchange of unpaired electrons (quasiparticle poisoning) \cite{Lei11,Bud11}.

\subsection{Read out}
\label{readout}

To read out a topological qubit one needs to remove the topological protection by coupling the Majorana fermions and then measure the quasiparticle parity. Tunnel coupling is one option, for example in the geometry of Fig.\ \ref{fig_QSHedge} the quasiparticle parity of two Majorana fermions can be measured by the difference in tunnel splitting \cite{Fu09}. The alternative Coulomb coupling allows the joint read out of more than a single qubit \cite{Has10b,Sau10b,Has11}. We concentrate on this second option, since two-qubit read out is required for quantum computations \cite{Nay08}.  

Consider a superconducting island (charge $Q$, superconducting phase $\phi$), containing $2N$ Majorana fermions. Cooper pairs can enter and leave the island via a Josephson junction. The read-out operation amounts to a measurement of $Q$ modulo $2e$. The conventional even-odd parity effect of the superconducting ground state does not apply here, because there is no energy cost of $\Delta$ for an unpaired electron in a midgap state. Indeed, the quasiparticle parity ${\cal P}$ does not enter in the Hamiltonian directly, but as a constraint on the eigenstates \cite{Fu10,Xu10},
\begin{equation}
\psi(\phi)=e^{i\pi{\cal P}}\psi(\phi+2\pi),\;\;{\cal P}=\tfrac{1}{2}+\tfrac{1}{2}i^{N}\gamma_{1}\gamma_{2}\cdots\gamma_{2N}.\label{Psiphi}
\end{equation}
This constraint enforces that the eigenvalues of the charge operator $Q=-2ei\,d/d\phi$ are even or odd multiples of $e$ when ${\cal P}$ equals $0$ or $1$, respectively.

\begin{figure}[tb]
\centerline{\includegraphics[width=0.7\linewidth]{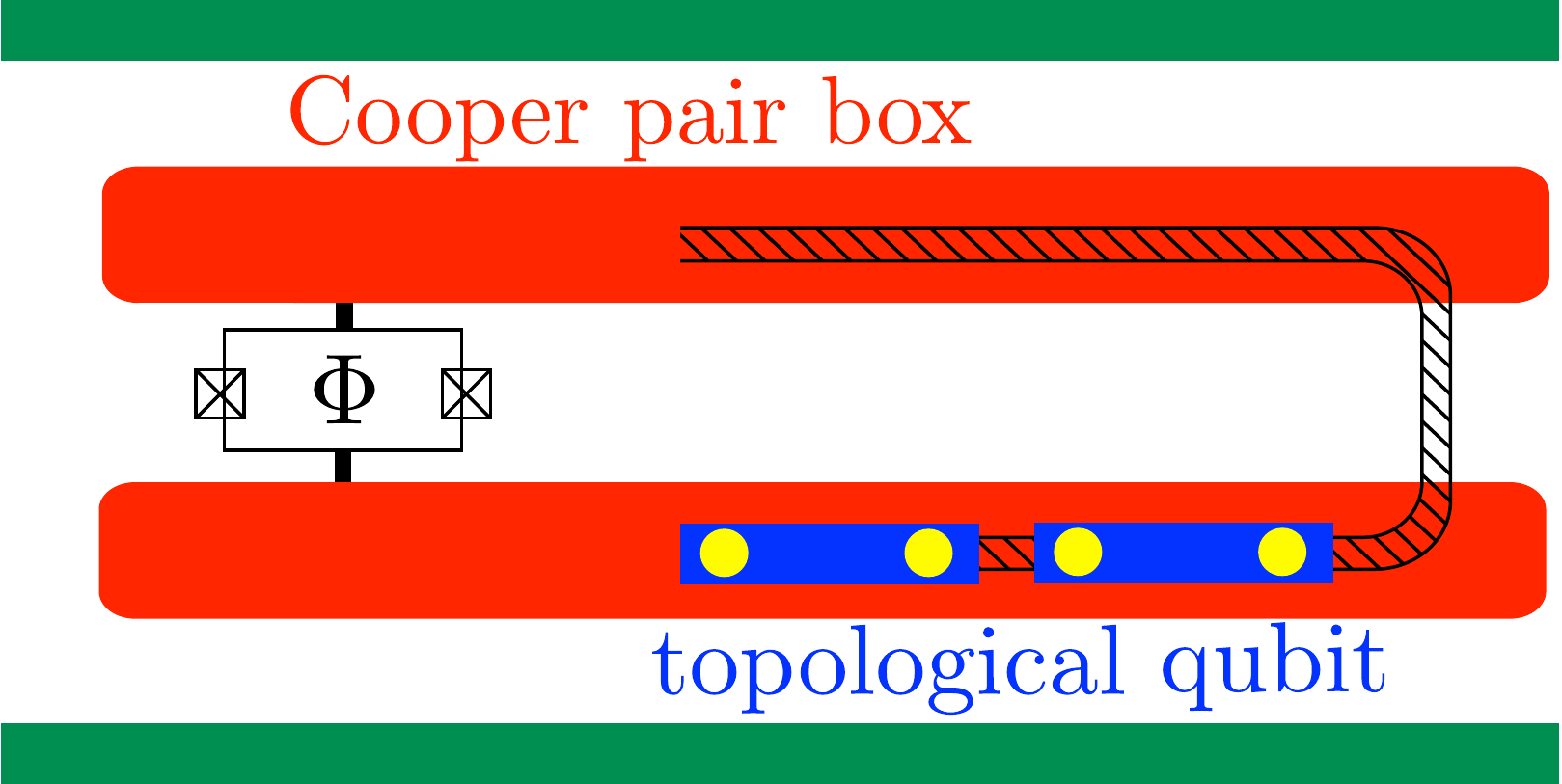}}
\caption{\label{fig_box}
Read out of a topological qubit in a Cooper pair box. Two superconducting islands (red), connected by a split Josephson junction (crosses) form the Cooper pair box. The topological qubit is formed by two pairs of Majorana fermions (yellow dots), at the end points of two undepleted segments (blue) of a semiconductor nanowire (shaded ribbon indicates the depleted region). A magnetic fiux $\Phi$ enclosed by the Josephson junction controls the charge sensitivity of the Cooper pair box. To read out the topological qubit, one pair of Majorana fermions is moved onto the other island. Depending on the quasiparticle parity, the resonance frequency in a superconducting transmission line enclosing the Cooper pair box (green) is shifted upwards or downwards by the amount given in Eq.\ \eqref{DeltaEdef}. Figure adapted from Ref.\ \onlinecite{Has11}.
}
\end{figure}

The parity constraint \eqref{Psiphi} modifies the energy of $\phi\mapsto\phi+2\pi$ quantum phase slips of the superconducting island, induced by the nonzero charging energy $E_{C}=e^{2}/2C$. The ${\cal P}$ dependence can be measured spectroscopically in a {\sc squid} geometry \cite{Has10b} or in a Cooper pair box \cite{Has11}. We show the latter geometry in Fig.\ \ref{fig_box}. 

The magnitude of the ${\cal  P}$-dependent energy shift $\Delta E$ in the Cooper pair box is exponentially sensitive to the ratio of Josephson and charging energies,
\begin{equation}
\Delta E= (2{\cal P}-1)U,\;\;U\simeq\sqrt{\hbar\omega_{p}E_{J}}e^{-\hbar\omega_{p}/E_{J}}.\label{DeltaEdef}
\end{equation}
(The frequency $\omega_{p}=\sqrt{8E_{C}E_{J}}/\hbar$ is the Josephson plasma frequency.) By varying the flux $\Phi$ through a split Josephson junction, the Josephson energy $E_{J}\propto\cos(e\Phi/\hbar)$ becomes tunable. In the transmon design of the Yale group, a variation of $E_{J}/E_{C}$ over two orders of magnitude has been realized \cite{Koc07}. The Coulomb coupling $U$ of the Majorana fermions can therefore be switched on and off by varying the flux.

\subsection{Braiding}
\label{sec_braiding}

In the two-dimensional geometry of Fig.\ \ref{fig_vortices} the Majorana bound states can be exchanged by moving the vortices around \cite{Fu08}. The Majorana fermions in the one-dimensional geometry of Fig.\ \ref{fig_box} are separated by insulating regions on a single nanowire, so they cannot be exchanged (at least not without rotating the wire itself \cite{Hal12}). The exchange of Majorana fermions, called ``braiding'', is needed to demonstrate their non-Abelian statistics \cite{Rea00}. It is also an essential ingredient of a topologically protected quantum computation \cite{Kit03}. In order to be able to exchange the Majoranas one can use a second nanowire, running parallel to the first and connected to it by side branches \cite{Ali11,Rom11}.

\begin{figure}[tb]
\centerline{\includegraphics[width=0.9\linewidth]{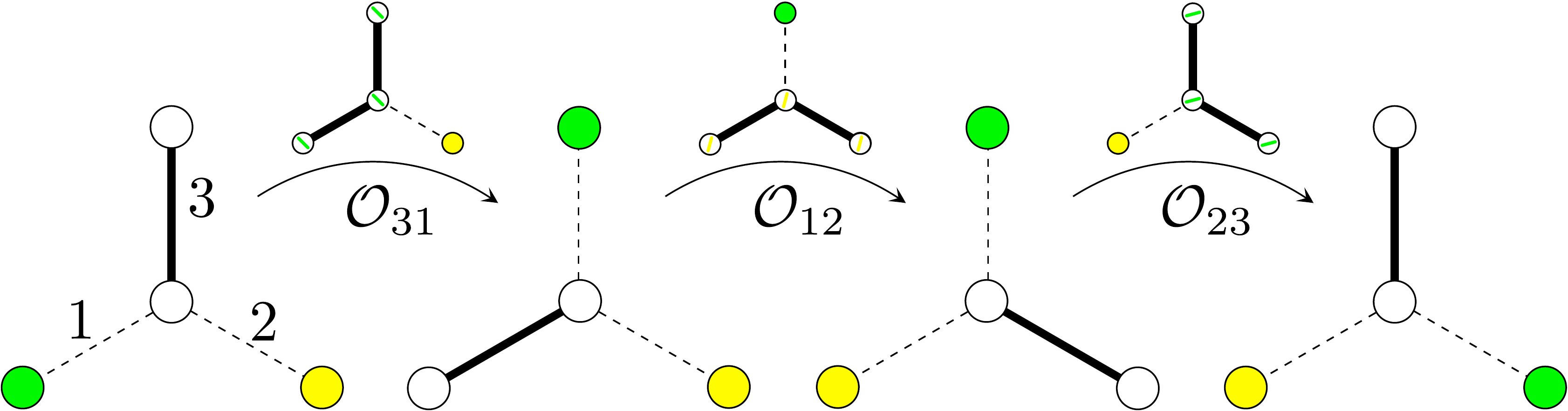}}\bigskip

\centerline{\includegraphics[width=0.7\linewidth]{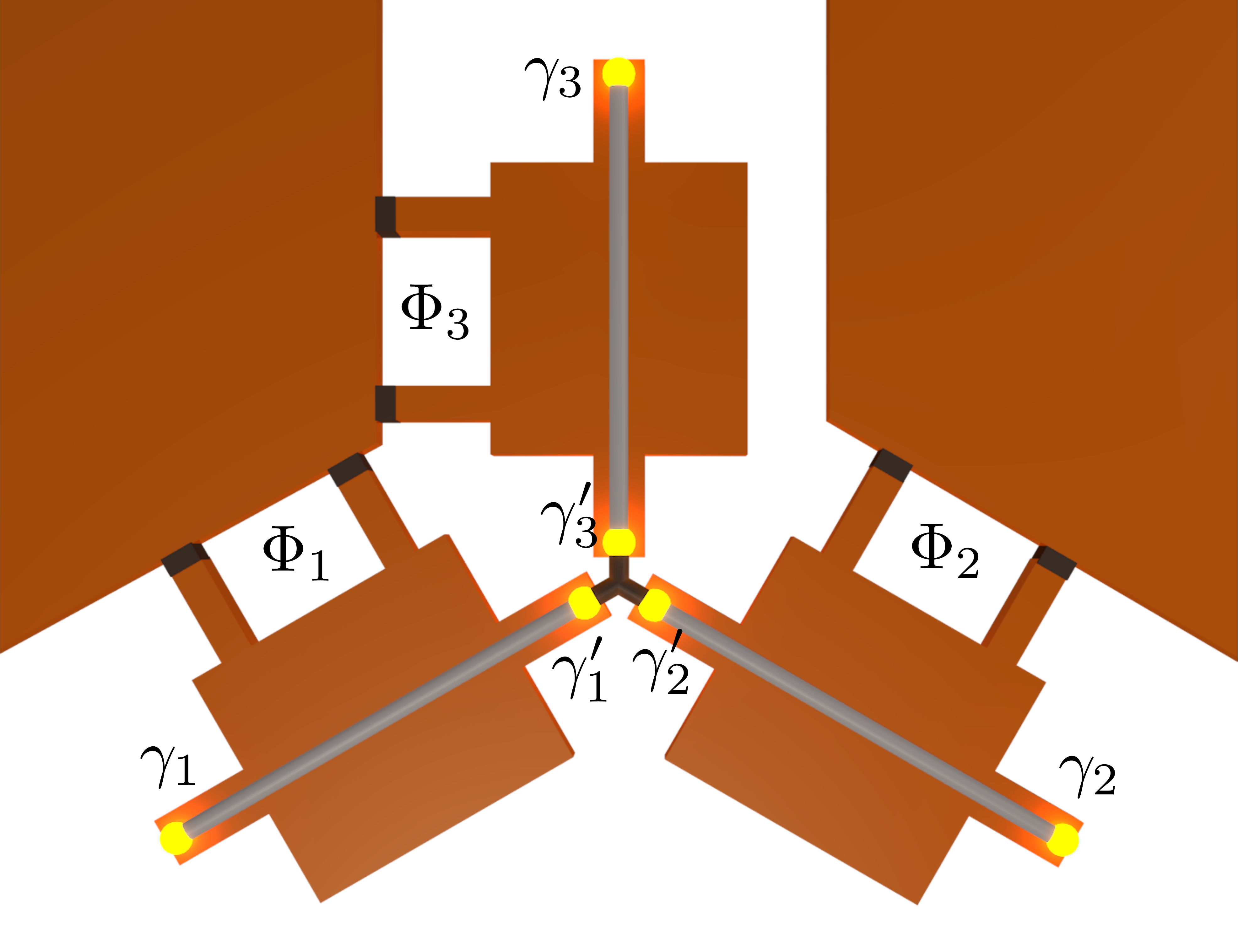}}
\caption{\label{fig_trijunction}
\textit{Lower panel:} Three Cooper pair boxes connected at a tri-junction via three overlapping Majorana fermions (which effectively produce a single zero-mode $\gamma_{0}=3^{-1/2}(\gamma'_{1}+\gamma'_{2}+\gamma'_{3})$ at the center).
\textit{Upper panel:} Schematic of the three steps of the braiding operation. The four Majoranas of the tri-junction (the three outer Majoranas $\gamma_{1},\gamma_{2},\gamma_{3}$ and the effective central Majorana $\gamma_{0}$) are represented by circles and the coupling $U_{k}$ is represented by lines (solid in the \textit{on} state, dashed in the \textit{off} state). White circles indicate strongly coupled Majoranas, colored circles those with a vanishingly small coupling. The small diagram above each arrow shows an intermediate stage, with one Majorana delocalized over three coupled sites. The three steps together exchange the Majoranas 1 and 2, which is a non-Abelian braiding operation.
Figure adapted from Ref.\ \onlinecite{Hec11b}.
}
\end{figure}

The minimal Hamiltonian that can describe the braiding contains three Majorana fermions $\gamma_{1},\gamma_{2},\gamma_{3}$ coupled to a fourth one $\gamma_{0}$,
\begin{equation}
H=\sum_{k=1}^{3}U_{k}i\gamma_{0}\gamma_{k}.\label{Hbraiding}
\end{equation}
The three parameters $U_{k}\geq 0$ can describe tunnel coupling \cite{Sau11} (tunable by a gate voltage) or Coulomb coupling \cite{Hec11b} (tunable by the flux through a Josephson junction). A tri-junction of three Cooper pair boxes that is described by this Hamiltonian is shown in Fig.\ \ref{fig_trijunction}. 

The braiding operation consists of three steps, denoted ${\cal O}_{31}$, ${\cal O}_{12}$, and ${\cal O}_{23}$. At the beginning and at the end of each step two of the couplings are \textit{off} and one coupling is \textit{on}. The step ${\cal O}_{kk'}$ consists of the sequence \textit{\{k,k'\,\}$\,=\,$\{on,off\,\}$\,\mapsto\,$\{on,on\}$\,\mapsto\,$\{off,on\}}. The effect of this sequence is to transfer the uncoupled Majorana $\gamma_{k'}\mapsto-\gamma_{k}$. (The minus sign appears in order to conserve the quasiparticle parity.) The result after the three steps shown in Fig.\ \ref{fig_trijunction} is that the Majoranas at sites 1 and 2 are switched, with a difference in sign, $\gamma_{2}\mapsto-\gamma_{1}$, $\gamma_{1}\mapsto\gamma_{2}$. The corresponding adiabatic time evolution operator in the Heisenberg representation $\gamma_{k}\mapsto{\cal U}\gamma_{k}{\cal U}^{\dagger}$ is given by
\begin{equation}
{\cal U}=\frac{1}{\sqrt 2}\bigl(1+\gamma_{1}\gamma_{2})=\exp\left(\frac{\pi}{4} \gamma_{1}\gamma_{2}\right)=\exp\left(i\frac{\pi}{4}\sigma_{z}\right).\label{U3T}
\end{equation}
This is the operator of Eq.\ \eqref{braiding}, representing a non-Abelian exchange operation.

\section{Outlook on the experimental progress}
\label{outlook}

As we have seen in Section \ref{howtomakethem}, there is no shortage of proposals for superconducting structures that should bind a Majorana zero-mode to a magnetic vortex or electrostatic defect. This gives much hope for a variety of experimental demonstrations in the coming years. There has already been a remarkable progress. 

A Josephson effect at the surface of a 3D topological insulator with superconducting electrodes has been observed in BiSb alloys \cite{Kas96}, and in crystalline $\text{Bi}_{2}\text{Se}_{3}$ \cite{Zha11,Wan12,Sac11,Wil12} and $\text{Bi}_{2}\text{Te}_{3}$ \cite{Vel12b,Qu12}. These experiments, and related Andreev conductance measurements\cite{Sas11,Kor11,Yan12,Wan12b}, all involve wide electrodes with a macroscopic number of occupied modes at the Fermi level. While the Josephson effect and the Andreev conductance show interesting and unusual features, these cannot be readily attributed to the single Majorana zero-mode (typically only one out of $10^{5}$ modes). Vortices (as in Fig.\ \ref{fig_vortices}), or other means of confinement would be needed to produce a Majorana bound state. 

The edge of a 2D topological insulator provides a single-mode conductor that could support spatially separated Majorana bound states, as in Fig.\ \ref{fig_QSHedge}. A superconducting proximity effect has been observed in an InAs/GaSb quantum well \cite{Kne11}, and also HgTe/CdTe would be a promising system --- if the Majorana can be confined to the superconducting interface by a magnetic insulator or magnetic field.

\begin{figure}[tb]
\includegraphics[width=0.8\linewidth]{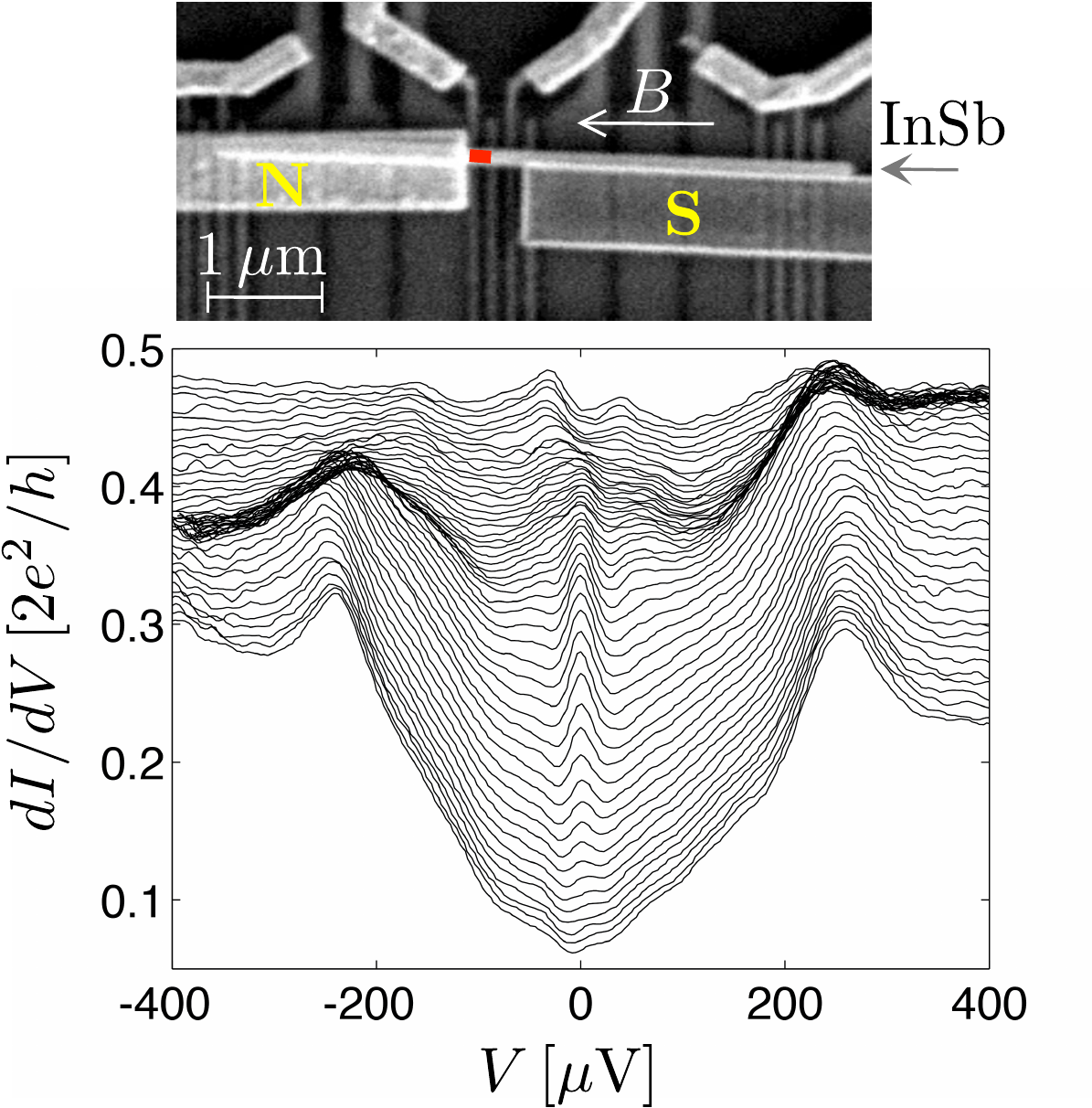}
\caption{\label{fig_InSb}
\textit{Top photograph:} InSb wire between a normal-metal (N) and a superconducting (S) electrode. A barrier gate creates a confined region (marked in red) at the interface with the superconductor. Other gates are used to locally vary the electron density. A magnetic field $B$ is applied parallel to the wire. \textit{Bottom graph:} Differential conductance at 60~mK for $B$ incrementing from 0 to 490~mT in 10~mT steps. (Traces are offset for clarity, except for the lowest trace at $B = 0$.) The peaks at $\pm 250\,\mu{\rm eV}$ correspond to the gap induced in the wire by the superconducting proximity effect. Upon increasing $B$ a peak develops at zero voltage, signaling the appearance of a Majorana zero-mode in the confined region. Figure adapted from Ref.\ \onlinecite{Mou12}.
}
\end{figure}

At this time of writing (April 2012), semiconductor nanowires have come furthest in the realization of Majorana fermions, following the proposal of Lutchyn et al.\ \cite{Lut10} and Oreg et al.\ \cite{Ore10}. Convincing evidence for a Majorana zero-mode in an InSb nanowire has been reported by Kouwenhoven and his group \cite{Mou12}, see Fig.\ \ref{fig_InSb}. These developments give hope that the rich variety of unusual properties of Majorana fermions, reviewed in this article, will soon be observed experimentally.\medskip

\noindent
{\bf Acknowledgments}\medskip\\
My own research on Majorana fermions was done in collaboration with A. R. Akhmerov, M. Burrello, T.-P. Choy, J. P. Dahlhaus, J. M. Edge, F. Hassler, B. van Heck, C.-Y. Hou, M. V. Medvedyeva, J. Nilsson, J. Tworzyd{\l}o, and M. Wimmer. Support by the Dutch Science Foundation NWO/FOM and by an ERC Advanced Investigator Grant is gratefully acknowledged.

\end{document}